%% file: main.tex
\documentclass[sigconf,10pt,nonacm]{acmart}

\AtBeginDocument{%
  }

\input{external/import}
\input{external/commands}

\begin{document}

\title[\system]{\system: Taming Step Misalignments in the Network \\ for Ring-based Collective Operations}

\author{Yuze Jin}
\email{jin.yuze@u.nus.edu}
\affiliation{%
  \institution{National University of Singapore}
  \country{Singapore}
}

\author{Xin Zhe Khooi}
\email{khooixz@u.nus.edu}
\affiliation{%
  \institution{National University of Singapore}
  \country{Singapore}
}

\author{Ruyi Yao}
\email{ryyao20@fudan.edu.cn}
\affiliation{%
  \institution{Fudan University}
  \country{China}
}

\author{Mun Choon Chan}
\email{dcscmc@nus.edu.sg}
\affiliation{%
  \institution{National University of Singapore}
  \country{Singapore}
}

\hyphenation{sym-pho-ny mis-align-ment syn-chro-ni-za-tion bottle-neck throu-gh-put work-loads im-prove-ments pro-to-type de-mon-stra-ting dis-tri-buted app-lica-tion inten-sive stra-gglers pipe-line pro-gress com-muni-ca-tion com-pu-ta-tion gen-era-tion ac-ce-le-ra-tors ten-ancy th-rou-gh in-gress e-gress with-in}

\settopmatter{printfolios=true, printccs=false, printacmref=false}

\input{sections/abstract}

\maketitle

\input{sections/introduction}
\input{sections/background}
\input{sections/motivation}
\input{sections/design}
\input{sections/evaluation}
\input{sections/discussion}
\input{sections/related_work}
\input{sections/conclusion}

\bibliographystyle{ACM-Reference-Format}
\balance
\bibliography{references}

\end{document}

%% file: external/import.tex
\usepackage[english]{babel}
\usepackage{xspace}
\usepackage[normalem]{ulem}
\useunder{\uline}{\ul}{}

\usepackage{amsmath}
\usepackage{amsfonts}
\usepackage{algorithmic}
\usepackage{graphicx}
\usepackage{textcomp}
\usepackage{xcolor}

\usepackage{tabularx,colortbl}
\newcolumntype{Y}{>{\centering\arraybackslash}X}

\usepackage[scaled=0.85]{beramono} 
\usepackage{tikz}
\usepackage{amsmath}
\usepackage{cleveref}
\usepackage{xspace}
\usepackage{epsfig}
\usepackage{svg}
\usepackage{subcaption} 
\usepackage{graphicx}
\usepackage{float}
\usepackage{upgreek}
\usepackage{algorithm}
\renewcommand{\algorithmiccomment}[1]{\bgroup\hfill//~#1\egroup}
\usepackage{caption}
\usepackage{url}
\usepackage{color}
\usepackage{multirow}
\usepackage{enumitem}
\usepackage{booktabs} 
\usepackage{makecell} 
\usepackage[binary-units=true]{siunitx}
\usepackage{balance}

%% file: external/commands.tex
\crefname{section}{§}{§§}
\crefname{section}{§}{§§}
\crefformat{section}{§#2#1#3}
\crefname{table}{Table}{Table}
\crefname{figure}{Fig.}{Fig.}
\crefname{listing}{Listing}{Listing}
\crefname{algorithm}{Alg.}{Alg.}
\crefname{objective}{\sc{\textbf{Objective}}}{\sc{\textbf{Objective}}}

\newcommand{\system}{\texttt{Symphony}\xspace}

\newcommand{\us}[1]{\SI{#1}{\micro\second}}
\newcommand{\ms}[1]{\SI{#1}{\milli\second}}
\newcommand{\second}[1]{\SI{#1}{\second}}
\newcommand{\byte}[1]{\SI{#1}{\byte}}
\newcommand{\kbyte}[1]{\SI{#1}{\kilo\byte}}

\newcommand{\Gbps}[1]{\SI{#1}{Gbps}}

\newcommand{\percent}[1]{\SI{#1}{\percent}}
\newcommand{\para}[1]{\vspace{1mm}{\noindent\textbf{#1.}}}

\newcommand{\parai}[1]{\vspace{1mm}\textit{#1}}

%% file: sections/abstract.tex
\begin{abstract}
Ring-based collective operations are widely used in distributed AI training due to
their efficient bandwidth utilization. 
While ring communication excels at pipelining,
its performance is heavily dependent on having synchronized step-wise progression.
This presents a mismatch to the underlying network conditions in practice: collective operations are 
vulnerable to network jitter and congestion,
leading to step misalignment and increased collective completion time. 
To that end, we propose \system{}, an in-network solution that detects pipeline step misalignment and mitigates its impact. 
\system{} introduces (1) a lightweight mechanism to track per-job pipeline progress and (2) a novel use of congestion signals to selectively throttle outpacing flows, allowing lagging flows to catch up without global coordination. 
Through simulations using Astra-Sim, we show that \system{} effectively mitigates step misalignments in ring-based collectives, resulting in up to 54\% improvement in job/collective communication time.
Finally, we prototype and validate \system{} on an Intel Tofino2 programmable switch to demonstrate its practicality.

\end{abstract}

%% file: sections/introduction.tex
\section{Introduction}
\label{sec:introduction}

\begin{figure*}[t!]
    \centering
    \includegraphics[width=0.9\linewidth]{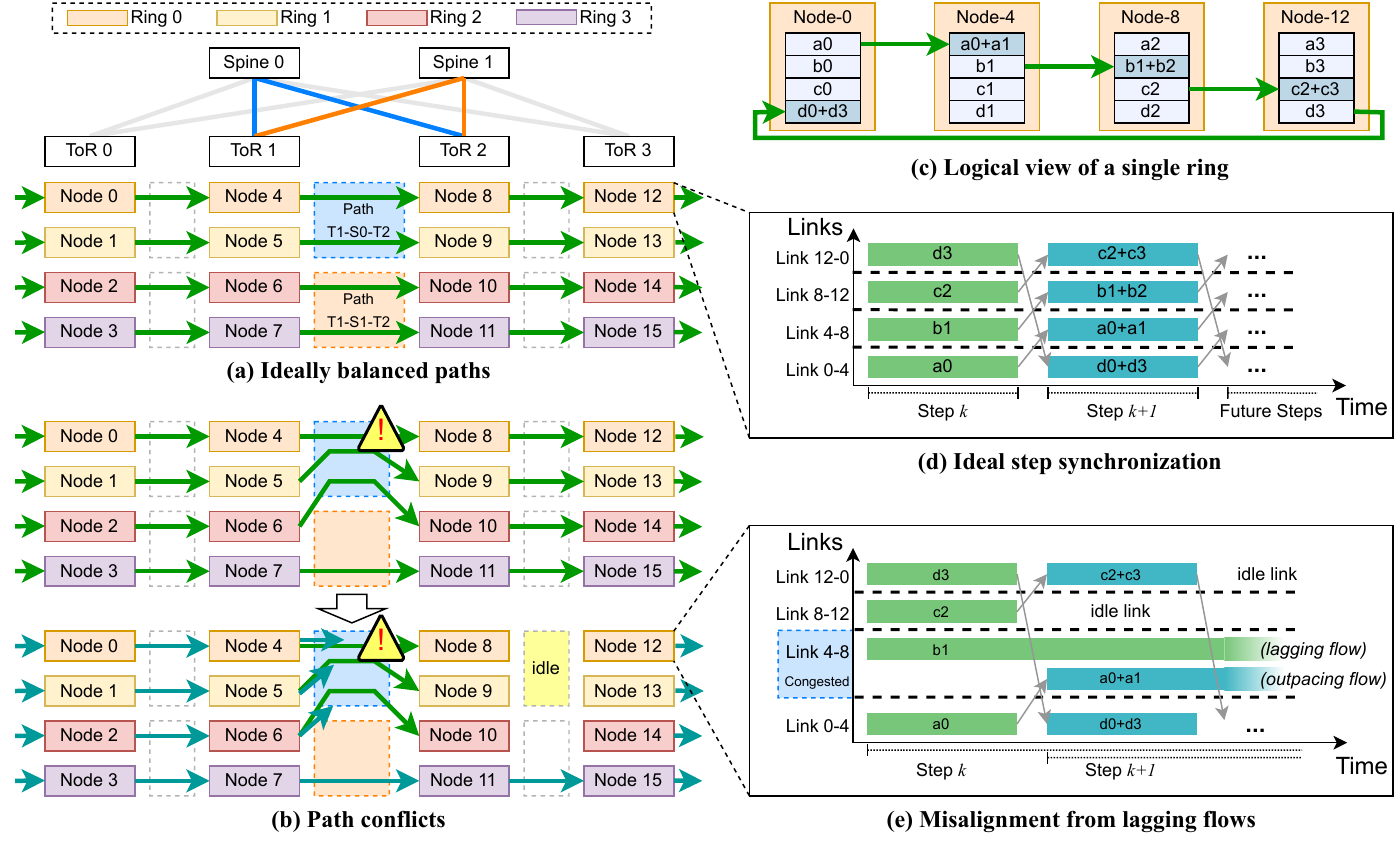}
    \caption{
    Ideal vs. Realistic behavior of ring‑based collective operations.
    (a–b) Even small path imbalances (e.g., ECMP hash collisions) cause uneven link sharing and flow accumulation, unlike the ideal case.
    (c) Logical view of a single ring.
    (d) Under ideal conditions, steps progress in perfect lockstep. 
    (e) In practice, congestion on a single link delays a step, causing subsequent steps to overlap and creating pipeline bubbles that amplify misalignment. 
    }
    \label{fig:motivation-example-one-view}
\end{figure*}

Training modern AI models, such as Large Language Models (LLMs), is resource-intensive and often spans weeks to months, requiring tens of thousands of GPUs running in parallel~\cite{2026-arxiv-meta-100kGPU-llama4, MegaScaleBytedance10000GPUsNSDI2024}. 
At this scale, training becomes communication-heavy: gradients and parameters must be exchanged across GPUs every iteration. 
In production clusters, communication bottlenecks often consume a significant portion of the total training iteration time, ranging from 16\% to over 50\% depending on the model size and parallelism strategy~\cite{dubey2024llama3herdmodels, 2024-SIGCOMM-MCCS, AlibabaHPNSIGCOMM2024}.

To fully utilize available bandwidth~\cite{2024-SC-gZCCL,2025-TACO-IBing}, today's large-scale distributed AI training and inference workloads~\cite{nccl,ncclrailoptimized,enhancedhorovod,sergeev2018horovodfasteasydistributed,baidu-allreduce,baidu-allreduce-gibiansky-2017} rely heavily on ring-based collective operations such as Ring-AllReduce.
This is because ring-based collective operations excel at pipelining data, where different parts of a message (chunks) are sent concurrently across different links. 
Through pipelining, not only can the available network bandwidth be fully utilized, but the latency of large messages can also be effectively amortized.

For ring-based collective operations to achieve their theoretical performance, they rely on the tight synchronization of the pipeline steps:
\emph{concurrent transfers between adjacent nodes must proceed in lockstep}.
Therefore, even minor network perturbations, such as ECMP hash collisions, transient congestion, physical-layer packet losses~\cite{2023-SIGCOMM-Protego, 2024-SIGCOMM-luberdma, 2025-SIGCOMM-DCP-RDMA-Lossy-Fabrics}, switch/link failures~\cite{MegaScaleBytedance10000GPUsNSDI2024,dubey2024llama3herdmodels}, or contention from background traffic in multi-tenant clusters~\cite{2024-SIGCOMM-MCCS}, can cause \emph{step misalignment}, where flows from different steps overlap on the same bottleneck link.
This overlap causes bandwidth contention, triggers cascading delays, and dramatically inflates collective completion time (CCT).

Unfortunately, today’s data center networks provide limited support for the synchronization needs of ring-based collectives.
Existing mechanisms optimize for aggregate throughput and link utilization through techniques like perfecting load balancing~\cite{2023-SIGCOMM-ConWeave,2017-SIGCOMM-DRILL,2014-SIGCOMM-Conga} and congestion control~\cite{2019-SIGCOMM-HPCC,2015-SIGCOMM-DCQCN}, to squeeze every bit of utilization out of the networking fabric.
However, such approaches blindly accelerate all flows, including those already ahead, and thus widen progress gaps and exacerbate misalignment.
Host‑side schedulers~\cite{2023-NSDI-TACCL, 2024-SIGCOMM-CRUX, 2024-SIGCOMM-MCCS} and application‑level straggler mitigation techniques~\cite{2016-SOCC-MLStraggler, 2017-ICML-Gradient-coding} also fall short because they lose visibility and control once packets enter the network fabric~\cite{2017-hotnets-gray-failure, MetaScaleSIGCOMM2024}.
In fact, stragglers remain an important issue for LLM training~\cite{MegaScaleBytedance10000GPUsNSDI2024, AlibabaHPNSIGCOMM2024, 2025-NSDI-optireduce}.

\textbf{\textit{Key insight:}}
Improving the performance of ring-based collective operations is not just about pushing more bandwidth; it requires the explicit control of the timing of flows so that steps stay synchronized as much as possible.

To this end, we propose \system{}, an in-network mechanism that detects and mitigates step misalignment for ring-based collective operations.
\system{} introduces a novel control strategy to restore pipeline alignment, where we throttle \emph{outpacing flows} to dynamically free up bandwidth for \emph{lagging flows} (or ``stragglers'') to catch up.
To achieve this, \system{} introduces two key ideas:
(1) \emph{progress tracking}: 
a lightweight per-job mechanism that infers real-time step progress using only packet metadata, and
(2) \emph{selective throttling}: 
a \emph{targeted} use of congestion signaling (i.e., ECN marking) to throttle {outpacing flows}, allowing {lagging flows} to catch up without explicit coordination.

Through extensive evaluations using Astra-Sim~\cite{won2023astrasim2,rashidi2020astrasim} with NS-3~\cite{ns3}, we show that \system{} effectively minimizes step alignments across various workloads, resulting in improved job and collective completion times (JCT/CCT) of up to 54\%.
To demonstrate \system{}'s practical deployability, we prototype \system{} on an Intel Tofino2~\cite{intel-tofino2} programmable switch, and show that \system{} can track step progress efficiently and throttle outpacing flows effectively to allow lagging flows to catch up.

\para{Contributions}
This paper is the first to identify and quantify step misalignment as a hidden bottleneck in ring‑based collective operations. 
Building on this insight, we introduce \system{}, a practical in‑network mechanism, validated with a Tofino2 hardware prototype, that detects and mitigates step misalignments through a novel use of ECN marking to realize the selective throttling of outpacing flows.

%% file: sections/background.tex
\begin{figure}
    \centering
    \includegraphics[width=0.85\linewidth]{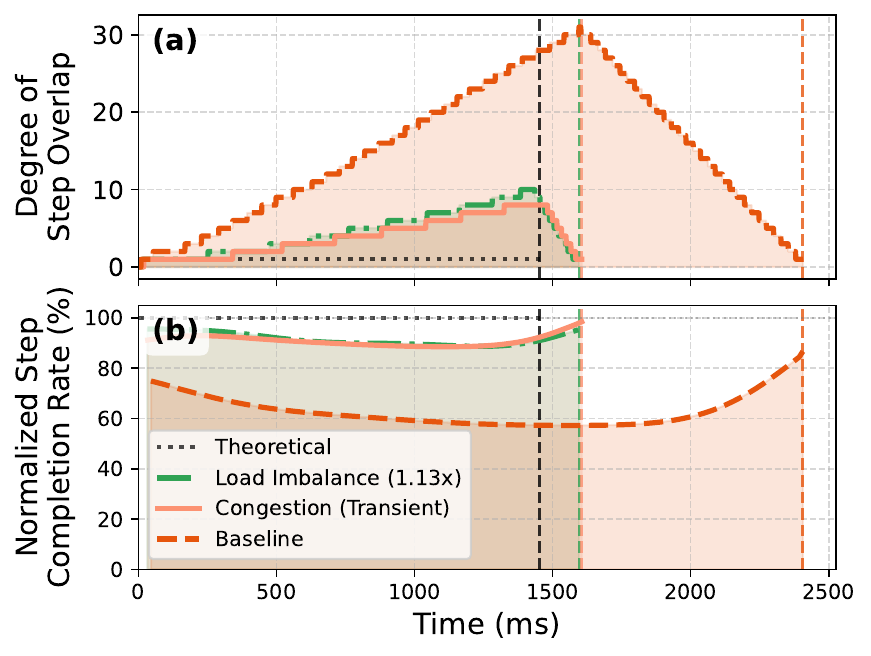}
    \vspace{-1em}
    \caption{
    Degree of step overlap and normalized step completion rates under different network conditions.
    }
    \vspace{-1em}
    \label{fig:motivation-example-experiments}
\end{figure}

\section{Background and Motivation}
\label{sec:bg-motivation}

\subsection{A primer on ring-based primitives}
\label{subsec:bg}

Ring-based algorithms are commonly adopted for
bandwidth-intensive LLM workloads due to their optimal link utilization via concurrent transmission~\cite{nccl, ncclrailoptimized, 2025-arxiv-DemystifyingNCCL}.
Taking Ring All-Reduce as an example, each of the $N$ nodes divides their data into $N$ equal-sized chunks and exchanges data chunks over $2(N-1)$ pipelined steps.
In each step, every node simultaneously transmits to its successor and receives from its predecessor, optionally undergoing computation (e.g., reduction) at each hop, as shown in \cref{fig:motivation-example-one-view}c.
Crucially, this throughput optimality relies on strict alignment: bandwidth is fully utilized only when concurrent flows start and finish in lockstep, leaving no links idle (\cref{fig:motivation-example-one-view}d).

In practice, modern AI clusters deploy high-performance hosts equipped with multiple GPUs and NICs (e.g., 8 GPUs/ NICs per host)~\cite{dubey2024llama3herdmodels, AlibabaHPNSIGCOMM2024, MegaScaleBytedance10000GPUsNSDI2024}.
To fully saturate the available bandwidth, the communication library does not run a single giant ring.
Instead, it establishes multiple parallel 1D rings (often referred to as channels or rails)~\cite{nccl, 2025-arxiv-DemystifyingNCCL, ncclrailoptimized, wang2024railonlylowcosthighperformancenetwork}. 
Each ring operates independently over a dedicated NIC port, pipelining chunks concurrently.
We illustrate an example in~\cref{fig:motivation-example-one-view}a.

%% file: sections/motivation.tex
\subsection{The problem of step misalignments}
\label{subsec:ring-non-ideal}

While the aligned progression described above is elegant in theory, it is fragile in practice.
Production networks face inherent runtime uncertainties, such as ECMP hash collisions~\cite{MetaScaleSIGCOMM2024, 2023-SIGCOMM-ConWeave} and transient congestion~\cite{MegaScaleBytedance10000GPUsNSDI2024, AlibabaHPNSIGCOMM2024}, physical-layer network failures~\cite{2023-SIGCOMM-Protego, dubey2024llama3herdmodels, 2024-SIGCOMM-luberdma, 2017-hotnets-gray-failure, 2025-SIGCOMM-DCP-RDMA-Lossy-Fabrics}, that inevitably disrupt synchronization.
We refer to this loss of lockstep progression as \emph{step misalignment}, where flows from different steps overlap on the same bottleneck link and fragment available bandwidth (see~\cref{fig:motivation-example-one-view}d).

A minor network perturbation is enough to trigger a cascading failure.
Consider the example shown in~\cref{fig:motivation-example-one-view}a, where a single 1D ring (e.g., the ring connecting Nodes 0, 4, 8, 12) within a larger collective job that spans multiple ToR switches.
In~\cref{fig:motivation-example-one-view}b, a load imbalance between ToR 1 and ToR 2 causes congestion on path T1-S0-T2.
Consequently, the flows from Node 4 to Node 8 (carrying step $k$) are impacted. 

This delay breaks the pipeline synchronization: Node 4 begins transmitting step $k+1$ before step $k$ completes. 
These concurrent steps now compete for the same bottleneck link, splitting bandwidth and further slowing down the already lagging step $k$.
This creates a destructive feedback loop: the worsening delay causes subsequent flows (step $k+2$ and beyond) to arrive and pile up. 
Hence, what started as a small link jitter cascades into significant performance degradation.

In practice, the completion times of flows within the same step can vary significantly due to runtime uncertainties and can persist even in well-provisioned networks~\cite{MetaScaleSIGCOMM2024, AlibabaHPNSIGCOMM2024, MegaScaleBytedance10000GPUsNSDI2024}, continuously disrupting alignment.

\para{Motivating example}
We show how this phenomenon manifests in~\cref{fig:motivation-example-experiments} using a multiple 1D Ring AllReduce workload (see~\cref{tab:simulation-parameters-table} in~\cref{sec:eval}). 
To pinpoint the sources of misalignment, we compare four scenarios:
(1) a \textit{Theoretical} lower bound assuming perfect lockstep;
(2) a \textit{Baseline} using standard ECMP routing;
and two controlled scenarios using static balanced routing with injected noise --
(3) \textit{Load Imbalance (1.13x)}, where we introduce a minor 1.13$\times$ traffic skew on a single hop, and
(4) \textit{Congestion (Transient)}, where we introduce light background traffic. 

\cref{fig:motivation-example-experiments}a shows that while the theoretical curve remains flat at 1, the baseline shows a runaway effect, with step overlap climbing to 30. 
Even under the scenarios with ``light'' perturbations, the degree of misalignment increases to 10 steps.
This accumulation of misalignment directly degrades the step completion rate (calculated as the inverse of inter-step completion intervals): \cref{fig:motivation-example-experiments}b shows that the normalized step completion rate drops significantly as overlap grows, inflating the CCT by 60\% in the baseline and 7\% even with minor perturbations.
These observations confirm a strong correlation between step alignments and the CCT.
Any ``loss'' of alignment, triggered by even minimal network jitter, is a fundamental bottleneck in ring-based communication.

\subsection{Existing approaches do not address the problem of step misalignments}
\label{subsec:baseline}

Here, we discuss why existing approaches fail to address step misalignments.

\para{Straggler mitigation in MLSys}
Traditional straggler mitigation techniques focus on compute‑side delays, arising from hardware heterogeneity, OS noise, or uneven workload distribution. 
These include redundancy‑based coding~\cite{2017-ICML-Gradient-coding, 2016-ISIT-straggler-codes, 2016-SOCC-MLStraggler}, dynamic scheduling, and compute‑overlap optimizations~\cite{2026-NSDI-straggler, 2022-eurosys-varuna, 2023-sosp-oobleck}. 
However, they treat the network as a black box and are unaware of the underlying collective operations.
Consequently, these solutions cannot address the problem of step misalignments.
\system{} complements these approaches by mitigating stragglers at the network-level.

\para{Framework-level communication scheduling}
Emerging training frameworks prioritize communication using DAG‑aware scheduling~\cite{2024-SIGCOMM-MCCS, 2023-NSDI-TACCL, alpaOSDI2022}, compression/fusion~\cite{2024-SIGCOMM-ccl-consumer-gradegpu, 2025-HPDC-SAFusion, MLSYS-2023-cupcake-compression}, and overlap~\cite{pipelineparallelism, pipedreamsSOSP2019, deepspeed} techniques. 
While these mechanisms optimize the logical execution order at the sender, they lose visibility the moment when data is enqueued at the NIC.
Runtime jitter inside the fabric, e.g., ECMP hash collisions or congestion, invalidates even perfect host‑side scheduling. 
Consequently, step misalignment persists even when the framework enforces ideal coordination at the endpoints, highlighting the need for an in‑network mechanism that manages the step-level progress.

\para{Network load balancing}
Data center load balancing solutions~\cite{cisco-silicon-one, 2017-SIGCOMM-DRILL,2023-SIGCOMM-ConWeave, 2016-SOSR-HULA, 2014-SIGCOMM-Conga} aim to distribute traffic as evenly as possible across paths to reduce congestion~\cite{MetaScaleSIGCOMM2024}.
Yet even ideal load balancing cannot prevent step misalignment, because unavoidable network jitter (e.g., due to gray failures~\cite{2017-hotnets-gray-failure,2023-SIGCOMM-Protego}) still causes temporal progress divergence among flows. 
Worse, LB is generally throughput‑centric, and it indiscriminately accelerates all flows, including those already ahead, thereby widening the relative progress gap that drives misalignment. 
Thus, LB solves spatial imbalance but does not address the temporal coupling of ring‑based primitives.

\para{Priority scheduling} 
One might consider prioritizing lagging flows directly inside switches~\cite{alizadeh2013pfabric, 2018-NSDI-SP-PIFO}. 
However, strict priority scheduling is unstable for RDMA traffic: line‑rate bursts rapidly fill shallow buffers~\cite{2019-SIGCOMM-HPCC, 2018-SIGCOMM-Homa}, causing severe starvation for non‑prioritized flows.
This, in turn, triggers aggressive congestion‑control responses (e.g., DCQCN backoff or PFC)~\cite{2015-SIGCOMM-DCQCN, 2015-SIGCOMM-TIMELY, 2016-SIGCOMM-PFC-RDMA}, which introduce oscillations and worsen tail latency. 
Instead of restoring alignment, this heavy‑handed approach often degrades overall CCT (more in~\cref{subsec:collective-comm}).

\parai{Key takeaway.}
Existing approaches fail because none of them coordinate the temporal progress of ring-based collective operations. 
They either optimize compute, logical scheduling, or spatial load distribution. 
None accounts for the fundamental constraint:
flows in a ring collective are tightly coupled, and their performance is dictated by the slowest step, not average throughput. 
This gap motivates the need for a new mechanism that explicitly manages step‑level progress in the network, which we introduce next.

%% file: sections/design.tex
\section{\system{}}
\label{sec:system}

\begin{figure}
    \centering
    \includegraphics[width=0.85\linewidth]{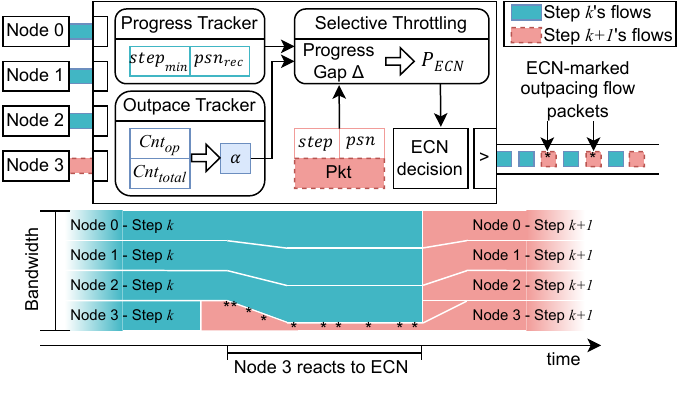}
    \vspace{-1em} 
    \caption{
    Overview of \system{}. 
    The switch performs progress tracking to identify and throttle outpacing flows, proactively freeing up bandwidth for lagging steps without global coordination. 
    }
    \label{fig:design-overview}
    \vspace{-1em} 
\end{figure}

\subsection{Overview of \system{}}
\label{subsec:overview}

\system{} addresses step misalignment by coordinating the temporal progress of flows directly inside the network. 
To achieve this, there are two key requirements:
\begin{itemize}[nosep,leftmargin=*]
    \item \textbf{R1:} 
    How to efficiently track the progress of collective pipelines across multiple jobs to enable timely detection of step misalignment?
    \item \textbf{R2:} 
    How 
    to selectively throttle \emph{outpacing flows} while yielding bandwidth to \emph{lagging flows}, thereby restoring pipeline alignment without compromising utilization?
 \end{itemize}

\system{} addresses step misalignment by combining two mechanisms: 
(1) \emph{progress tracking} (\cref{subsec:dataplane-design}): lightweight per‑job tracking that infers real‑time progress of each step from packet metadata, while maintaining minimum states, and
(2) \emph{selective throttling} (\cref{subsec:straggler-mitigation}): selective throttling that uses ECN to apply backpressure only to outpacing flows, allowing lagging flows to catch up.

As illustrated in \cref{fig:design-overview}, when a flow from step $k+1$ begins overlapping with a still‑unfinished step $k$, \system{} increases the marking probability of the outpacing flow. 
This slows the outpacing flow at the source via the existing congestion‑control loop (e.g., DCQCN), implicitly reallocating bandwidth to the lagging step and restoring alignment.

\subsection{Problem modeling and assumptions}
\label{subsec:problem-modeling}

\subsubsection{Problem modeling}
We model the network as a set of switches serving 
concurrent collective jobs.
\system{} relies on the following abstractions to perform alignment.

\para{Traffic granularity} 
We conceptualize the collective traffic hierarchy in three levels:
\begin{itemize}[leftmargin=*,nosep]
    \item \textbf{Job}: 
    A distributed training task consisting of multiple collective operations. 
    We assume a multi-tenant DC environment where multiple jobs may coexist concurrently.
    
    \item \textbf{Step ($s$)}: 
    A logical stage in the ring-based collective (e.g., Ring AllReduce). A collective operation proceeds in a sequence of synchronized steps $s_0, s_1, ..., s_n$.
    
    \item \textbf{Flow ($f$)}: 
    The data transmission required for a single node to complete one step. Ideally, all flows in step $s_k$ should complete before any flow in step $s_{k+1}$ begins. 
    We assume that for a given step, the data volume to be transferred is uniform across nodes (typical for ring-based collectives).
\end{itemize}

\para{Switch visibility}
We assume the switch can parse two critical metadata fields from each packet's header.
\begin{itemize}[leftmargin=*,nosep]
    \item \textbf{Step index} ($step$):
    Indicates the step to which the packet’s flow belongs.  
    Comparing these indices between flows allows \system{} to determine the degree of step misalignment between outpacing and lagging flows.

    \item \textbf{Packet sequence number} ($psn$): 
    Denotes the packet's position within its flow.
    This provides a fine-grained estimate of data transferred by that flow, allowing \system{} to determine the progress of a flow within its step.
\end{itemize}

\para{Objective}
The goal of \system{} is to minimize the misalignment gap between the fastest outpacing flows and the slowest lagging flows at bottleneck link(s), 
by giving a larger share of bandwidth to lagging flow(s).
At the same time, 
we target a fully distributed solution: each switch minimizes misalignment based solely on its local visibility of the traffic, eliminating the need for coordination.

\subsubsection{Technical assumptions}
\system{} assumes the ability to distinguish traffic at both job and step granularity.

While such metadata is not natively encoded in legacy RDMA fabrics, \system{} aligns with emerging AI/HPC interconnect standards such as the Ultra Ethernet Consortium (UEC)~\cite{uec-specification-2025}. 
Specifically, the \emph{JobID}~\cite[Table 2-39]{uec-specification-2025} and \emph{PDCID} (Packet Delivery Context ID)~\cite[Table 3-33]{uec-specification-2025} fields in the UEC specification map directly to \system{}'s job identifier and step index, respectively~\cite{uec-specification-2025}. 
In our RDMA-based testbed, we emulate this capability by embedding the job ID and step index into the RDMA reserved fields and UDP source port (details in \cref{subsec:testbed-eval}).

With these assumptions in place, we now detail the design of \system{}.
While our model accounts for multiple concurrent jobs, the core logic of \system{} operates independently per job on a switch. 
For simplicity, we will present \system{} running with a single job in \cref{subsec:straggler-mitigation}--\cref{subsec:dataplane-design}.
Then we discuss support for multi-tenancy in \cref{subsec:design-multi-tenant}.

\subsection{Selective throttling}
\label{subsec:straggler-mitigation}

To efficiently track misalignment across collective steps, each switch maintains a compact \emph{Per-Job State Block}.
This design ensures scalability with constant memory overhead per job. The state block comprises two logical components:

\para{1. Progress Tracking State} 
These variables serve as the synchronization anchors:
\begin{itemize}[leftmargin=*]
    \item $step_{min}$: 
    The smallest step index currently observed among all active flows. 
    This serves as a coarse-grained indicator of the collective's global synchronization anchor.
    
    \item $psn_{rec}$: 
    The estimated packet sequence number (PSN) represents the progress of lagging flows within $step_{min}$.
    This serves as a fine-grained reference for intra-step progress (approximation details in \cref{subsec:dataplane-design}).
\end{itemize}

\para{2. Adaptive Control State} 
To enable dynamic throttling, \system{} maintains a scalar state $\alpha(t)$ and transient counters to monitor traffic intensity:
\begin{itemize}[leftmargin=*]
    \item $\alpha(t)$: The adaptive aggressiveness factor. 
    This factor calibrates the calculated progress gap    
    based on the historical persistence of misalignment.
    
    \item $\rho(t)$: The misalignment intensity (transient), representing the fraction of outpacing traffic observed in the current time window $t$.
    
    \item $\tau$: A static tolerance threshold for $\rho(t)$, determining the 
    condition 
    for increasing $\alpha(t)$.
\end{itemize}

\para{Detection strategy}
\system{} examines each dequeued packet $i$ by comparing its progress ($step_{i}$, $psn_{i}$) against the global lagging flow reference ($step_{min}$, $psn_{rec}$).
If the packet belongs to a lagging 
flow ($step_{i} \le step_{min}$), no action is needed. 
However, if the packet belongs to an outpacing flow ($step_{i} > step_{min}$), \system{} calculates a \textit{Progress Gap}, denoted as $\Delta$, to quantify the severity of misalignment.

\begin{equation}
\Delta(t) =\alpha(t) \cdot \frac{psn_{i}}{psn_{rec}}
\label{eqn:progress-gap}
\end{equation}

$\Delta$ has two components. The first component
$\frac{psn_{i}}{psn_{rec}}$ captures difference in flow-level progress for the current packet.
On the other hand, the second component \textit{adaptive aggressiveness factor} $\alpha(t)$ captures the misalignment observed over time. 
\system{} updates $\alpha(t)$ by accumulating the observed misalignment over time as follows:

\begin{equation} 
\alpha(t) = \max\left(1, \alpha(t-1) + \delta(t)\right) 
\label{eqn:alpha-update} 
\end{equation}

\begin{equation} 
\delta(t) = \begin{cases} 
+1, & \text{if } \rho(t) \ge \tau \\
-1, & \text{if } \rho(t) < \tau
\end{cases} 
\label{eqn:delta-def} 
\end{equation}

Here, $\rho(t)$ represents the ratio of outpacing traffic volume (packets from steps $> step_{min}$) observed in the current time window, and $\tau$ is a small constant (e.g., 0.25).
$\delta$ allows \system{} to differentiate between \textit{transient} and \textit{persistent} misalignment. 
With persistent misalignment, $\alpha(t)$ increases, which in turn scales the throttling probability when outpacing flows dominate the bandwidth for extended periods.
Conversely, when misalignment reduces ($\rho < \tau$), $\alpha(t)$ decreases correspondingly.

\para{Probabilistic throttling}
Based on the calculated gap, \linebreak \system{} applies ECN marking with a probability proportional to $\Delta$ to enforce throttling.
The marking probability is formulated as:
\begin{equation}
    P_{ECN}(\Delta) = \min(1, k \cdot \Delta)
    \label{eqn:ecn-marking}
\end{equation}
where $k$ is a control parameter that determines the gain of the throttling feedback loop.
Since \system relies on relative progress rather than absolute traffic volume, an appropriate $k$ is primarily determined by the control loop latency (i.e., network RTT) rather than specific workload characteristics.
We demonstrate in~\cref{subsubsec:robustness} that a static, robust $k$ suffices for diverse workloads.

\para{Coexistence with congestion control mechanisms}
It is important to note that \system{} is designed to complement, not replace, existing congestion control mechanisms (e.g., DCQCN~\cite{2015-SIGCOMM-DCQCN}). 
\system{} operates in a \emph{logical \emph{OR} relationship} with the underlying congestion control: a packet is marked if \emph{either} the switch buffer exceeds the standard ECN threshold (indicating physical congestion) or \system{}'s logic triggers a mark based on~\cref{eqn:ecn-marking} (indicating logical misalignment). 
This design allows \system{} to mitigate step misalignment even when the network is not physically congested, while retaining the fail-safe protection of traditional mechanisms.

As summarized in \cref{alg:ecn_marking_algo}, if $step_{i} \le step_{min}$ or the lagging progress is insufficient ($psn_{rec} \le N_{warmup}$), the packet is treated as normal flow (skipping \system{} marking); otherwise, it is subject to the selective throttling described above. 
This warm-up threshold ensures numerical stability during the initial transient phase of a step.

\begin{algorithm}[t]
\caption{\system{} Selective Throttling Logic}
\label{alg:ecn_marking_algo}

\begin{minipage}[t]{0.48\linewidth}
    \textbf{Inputs:} $step$, $psn$ \\
    \textbf{Output:} $to\_mark\_ecn$ \\
    \textbf{Parameter:} $k$
\end{minipage}
\hfill
\begin{minipage}[t]{0.48\linewidth}
    \textbf{Constant:} $N_{warmup}$ \\
    \textbf{States:} $step_{\min}$, $psn_{\mathrm{rec}}$, $\alpha$
\end{minipage}

\vspace{2pt}
\hrule

\begin{algorithmic}[1]
    \STATE \textbf{Function} \textsc{Symphony}($step$, $psn$)
    
    \STATE \textsc{UpdateTrafficStats}($step$) \textit{// Updates $Cnt_{total}$, $Cnt_{op}$}
    
    \IF{\textsc{Is\_RDMA\_LastWrite}()}
        \STATE $step_{\min} \gets step + 1$
        \STATE $psn_{\mathrm{rec}} \gets 0$ 
    \ELSIF{$step < step_{\min}$}
        \STATE $step_{\min}, psn_{\mathrm{rec}} \gets step, psn$
    \ELSIF{$step = step_{\min}$}
        \STATE $psn_{\mathrm{rec}} \gets \max(psn_{\mathrm{rec}}, psn)$
    \ENDIF
    \IF{$(step \le step_{\min}) \OR (psn_{rec} \le N_{warmup})$}
        \STATE $to\_mark\_ecn \gets false$
    \ELSE
        \STATE $\Delta \gets \alpha \times (psn / psn_{\mathrm{rec}})$ \textit{// $\alpha$ is updated periodically}
        \STATE $P_{\Delta} \gets \min(1,\; k \cdot \Delta)$
        \STATE $to\_mark\_ecn \gets \textsc{TossCoin}(P_{\Delta})$
    \ENDIF
    \RETURN $to\_mark\_ecn$
\end{algorithmic}
\end{algorithm}

Next, we detail how \system{} efficiently approximates the state variables ($step_{min}$ and $psn_{rec}$) in the data plane to ensure hardware feasibility and robustness.

\subsection{Progress tracking}
\label{subsec:dataplane-design}

\subsubsection{Tracking inter-step progress}
\label{subsec:step-monitoring}
To track the global minimum step $step_{min}$ within the hardware constraints of modern network switches~\cite{2013-SIGCOMM-RMT}, \system{} employs an \textit{optimistic advancement} strategy with \textit{lazy correction}.

\para{State maintenance}
\system{} monitors transport headers for step completion signals (e.g., the ``LAST'' bit in RDMA WRITEs)\footnote{For network transports other than RDMA, similar semantics are also present~\cite{uec-specification-2025}, and \system{} can be adapted accordingly.}. 
Upon observing a packet with the ``LAST'' bit set for $step_{i}$, the switch tentatively updates $step_{min} \leftarrow step_{i+1}$.
This assignment operates on the optimistic assumption that the collective is progressing uniformly.

\para{Self-correction and robustness}
A critical challenge is maintaining state consistency in an unreliable network environment. 
\system{} addresses these edge cases through a series of resilient self-correcting logic:
\begin{itemize}[leftmargin=*]
    \item \textbf{Packet reordering \& late arrivals:} 
    If the switch has advanced $step_{min}$ to $step_{i+1}$ (due to an outpacing flow) but subsequently receives a packet from the previous $step_{i}$ (e.g., a severely lagging flow), the comparison $step_{i} < step_{min}$ triggers an immediate correction. 
    The switch updates $step_{min} \leftarrow step_{i}$, ensuring that the throttling logic correctly reflects the earlier system state.
    Furthermore, since \system{} relies on value assignment rather than incrementing counters, duplicate packets (e.g., retransmissions) are idempotent and do not corrupt the state.

    \item \textbf{Packet loss (e.g., missed ``LAST'' bit):} 
    If the packet carrying the ``LAST'' bit is dropped (e.g., due to transient link failures~\cite{2023-SIGCOMM-Protego}), the switch fails to advance $step_{min}$. 
    In such a case, incoming packets from the subsequent step ($step_{i+1}$) are identified as \emph{outpacing} relative to the stale $step_{min}$, triggering throttling. 
    This behavior acts as a natural \textit{fail-safe}: by suppressing the leading wave of traffic, 
    \system{} mitigates contention, thereby facilitating the eventual retransmission of the lost packet in $step_{i}$.
    Since flows belonging to the (stale) $step_{min}$ are never throttled, the retransmission faces minimal contention, ensuring rapid convergence once the transport reliability mechanism (e.g., Go-Back-N) succeeds.
\end{itemize}

Finally, given the high packet rate of RDMA, any transient state inconsistency is corrected within microseconds by subsequent arriving packets, ensuring negligible impact on throttling accuracy.

\subsubsection{Monitoring intra-step progress}
\label{subsec:psn-monitoring}

Ideally, to precisely quantify the misalignment gap $\Delta$, \system{} would calculate the ratio between the current flow's progress and that of the \textit{slowest} flow (the ``tail'') in the lagging step ($step_{min}$).
However, tracking the true global minimum PSN requires maintaining per-flow counters to compare all active flows, which incurs a memory overhead of $O(N)$ that scales linearly with the number of concurrent flows. 
This is prohibitively expensive due to switch hardware constraints~\cite{2021-NetSoft-dSketch}.

\para{Conservative approximation with max PSN}
To ensure scalability, we adopt a constant-state approximation. 
Instead of tracking the slowest flow, \system{} tracks the \emph{maximum} PSN observed among flows in the current $step_{min}$, denoted as $psn_{rec}$. 
This approach requires only keeping track of a single state per job.

We acknowledge that using the maximum PSN is a \textit{conservative} estimate. 
By defining the progress of the lagging step ($psn_{rec}$) based on the ``head'' of the lagging pack rather than the ``tail'', we effectively increase the denominator in the progress gap calculation (~\cref{eqn:ecn-marking}). 
This yields a smaller $\Delta$ and, consequently, a more lenient throttling probability.
This design choice is intentional: it acts as a safeguard against over-throttling, ensuring that \system{} intervenes only when the outpacing flow is substantially ahead of the fastest flow in the lagging step.

\para{Time-windowed estimation}
A potential side effect of tracking the maximum PSN is ``staleness'': a single bursty flow in $step_{min}$ could set a high $psn_{rec}$ early on, which would then persist and mask the presence of slower flows, preventing necessary throttling.

To mitigate this, we employ a \textit{time-windowed max} mechanism. 
The switch resets the $psn_{rec}$ register at periodic intervals (e.g., every \us{100} in our implementation).
Within each interval, $psn_{rec}$ tracks the maximum PSN of \emph{currently active} packets among $step_{min}$. 
This ensures that $psn_{rec}$ remains a timely indicator of the effective throughput of the lagging step, rather than a historical high-water mark.

Our evaluation (\cref{sec:eval}) confirms that this approximation faithfully captures flow dynamics and enables effective throttling without per-flow book-keeping.

\subsubsection{Monitoring outpacing traffic ratio}
\label{subsubsec:outpacing-traffic-ratio}

To periodically update $\alpha(t)$, \system{} must quantify the misalignment intensity $\rho(t)$. Updating $\alpha(t)$ on a per-packet basis would be susceptible to micro-burst noise and computationally expensive.
\system{} addresses these using a window-based aggregation mechanism.

\para{Windowed Aggregation}
\system{} maintains two 
counters per job:
a total packet counter $Cnt_{total}$ and an outpacing packet counter $Cnt_{op}$. 
These counters accumulate traffic statistics over a discrete time window $T_{win}$ (e.g., $100\mu s$). 

\para{Update Logic}
At the end of each window, \system{} updates $\alpha(t)$.
Rather than computing the exact floating-point ratio $\rho(t) = Cnt_{op} / Cnt_{total}$, \system{} simply checks if the outpacing traffic exceeds the tolerance threshold:
\begin{equation}
Cnt_{op} \ge \tau \cdot Cnt_{total}
\label{eqn:outpacing-check}
\end{equation}
If the condition holds, it implies $\rho(t) \ge \tau$, triggering an increment in $\alpha(t)$; otherwise, $\alpha(t)$ decays. 
This integer-based comparison is HW-friendly and avoids complex division operations. 
Finally, to ensure statistical significance, the update is skipped if the total sample count $C_{total}$ is insufficient.

\para{Handling Noise and Transience} 
To ensure control loop stability, \system{} incorporates two safeguards. 
First, to prevent oscillation during low-traffic periods, we apply a \textit{Sample Guard}: the update logic is triggered only if the window's sample count exceeds a minimum threshold ($Cnt_{total} > N_{sample}$). 
Second, the integral nature of $\alpha(t)$ naturally dampens transient inconsistencies (e.g., mismatches during $step_{min}$ transitions), as isolated ``noisy'' windows have negligible impact on the cumulative system state.

\subsection{Supporting multi-tenancy}
\label{subsec:design-multi-tenant}

\para{Per-job state isolation} 
To accommodate concurrent jobs without interference, \system{} maintains isolated progress tracking states and counters
for each job, indexed by a unique identifier, \texttt{Job\_ID}.
Upon packet arrival, the switch parser extracts the \texttt{Job\_ID} and utilizes it to index the corresponding state slot in the register memory.
This design ensures that the control loop of one job operates completely independently from others, effectively instantiating a dedicated ``virtual regulator'' for each tenant.

\para{Control plane orchestration} 
Identity management integrates naturally with the SDN control plane. 
When a new training job is scheduled, the cluster scheduler requests a \texttt{Job\_ID} from the \system{} controller. 
The controller then programs the necessary match-action rules onto the relevant switches along the job's path. 
This centralized coordination ensures ID uniqueness and facilitates policy-based resource allocation, enabling \system{} to seamlessly integrate into existing cluster management stacks.

\para{Scalability}
\system{} is lightweight because it tracks progress at the coarse granularity of steps and aggregates intra-step markers; rather than maintaining per-flow sequence numbers, the memory footprint per job is minimal.
This efficiency allows \system{} to scale to support thousands of concurrent jobs within the on-chip SRAM capacity constraints of commodity programmable switches. 
For instance, even when scaled to support 16k concurrent jobs, the total state memory footprint of \system{} remains in the order of hundreds of KB. This overhead is negligible for modern switching ASICs, which typically contain tens of MB of SRAM~\cite{2013-SIGCOMM-RMT}.

\subsection{Putting things together}
\label{subsec:putting-things-together}

By integrating the progress tracking and throttling mechanisms described above, \system{} delivers a cohesive solution that offers two advantages.

First, \system{} enables selective throttling to dynamically rebalance bandwidth: outpacing steps are modulated to yield bandwidth resources that lagging flows can opportunistically utilize. 
Second, by leveraging pre-existing congestion signals (\cref{subsec:straggler-mitigation}) and lightweight progress tracking (\cref{subsec:dataplane-design}), \system{} is designed to be deployable on existing systems. 
We discuss data plane implementation challenges, including arithmetic limits on programmable switches, later in~\cref{subsec:testbed-eval}.

%% file: sections/evaluation.tex
\section{Evaluation}
\label{sec:eval}

We evaluate \system using both packet-level simulations and hardware prototyping. 
Our evaluation seeks to answer the following questions:

\begin{itemize}[leftmargin=*]
    \item Can \system effectively mitigate step misalignment and restore pipeline synchronization? (\cref{subsubsec:reduce-misalignment})

    \item Does \system improve collective communication performance and end-to-end AI training? (\cref{subsec:collective-comm} - \cref{subsubsec:end-to-end})
    
    \item Is \system effective in multi-tenant environments? (\cref{subsubsec:multi-tenant})
    
    \item Is \system robust across varying network configurations? (\cref{subsubsec:robustness})
    
\end{itemize}
Finally, we validate the feasibility of \system using a \linebreak Tofino2-based hardware prototype (\cref{subsec:testbed-eval}).

\subsection{Simulation setup}
\label{subsec:software-simulations}

\begin{table}
\centering
\caption{Default parameters for software simulations.}
\vspace{-1em} 
\label{tab:simulation-parameters-table}
\renewcommand{\arraystretch}{0.9} 
\resizebox{0.9\linewidth}{!}{%
    \begin{tabular}{c|p{0.6\linewidth}|l} 
    \hline
    \textbf{Param} & \textbf{Definition} & \textbf{Default Values} \\ 
    \hline
    \hline

    $k$ & Selective throttling parameter & 0.01 \\ \hline

    $P$ & Physical topology &
    \makecell[l]{4 ToR, 4 Spine \\ 2/4 Core switches} \\ \hline

    $S$ & Network Scale & 32 nodes \\ \hline

    $L$ & Logical topology (2D ring dim.) &
    \makecell[l]{$8 \times 4$ (32 nodes) \\ $32 \times 4$ (128 nodes)} \\ \hline

    $S_{\text{chunk}}$ & \rule{0pt}{2.5ex}Chunk size ($\frac{\text{collective size}}{\text{\# of nodes}}$) & \makecell[l]{8 MB} \\[1ex] \hline

    \end{tabular}%
}
\vspace{-1em} 
\end{table}

We use software simulations to study how \system{} can improve the communication performance for various workloads at scale.
We use Astra-Sim~\cite{rashidi2020astrasim, won2023astrasim2} with NS-3~\cite{ns3} to perform the simulations, where Astra-Sim models collective operations and pipelining behavior, while NS-3 simulates the underlying network stack.

\para{Implementation}
The core logic of \system is implemented using $\approx$150 lines of code in NS-3's memory management unit (MMU) to perform both progress tracking and selective throttling.

\para{Network topology}
We simulate a standard two-tier leaf-spine topology with \Gbps{10} links. 
Clusters of up to 128 nodes reside within a single spine group, while larger clusters (256/512 nodes) employ multi-pod interconnects with oversubscription ratios of 1:2 to 1:8. 
The underlying network uses ECMP with 5-tuple flow hashing for load balancing. 

\para{Simulation parameters}
We follow the RDMA configuration of HPCC~\cite{2019-SIGCOMM-HPCC}. 
\system operates alongside DCQCN.
We utilize standard RED parameters ($K_{min}=50, K_{max}=100, P_{max}=0.2$)~\cite{2019-SIGCOMM-HPCC} without specific tuning\footnote{
This choice is deliberate: \system addresses \textit{temporal misalignment} (logical coordination), which is orthogonal to the physical \textit{rate mismatches} handled by CC. 
Even a perfectly tuned CC cannot identify that a non-congested fast flow should yield for synchronization. 
Thus, using standard parameters confirms that our gains stem from structural alignment rather than parameter sensitivity.}.
The control loop parameters are fixed at $\tau=0.25$ and $T_{win}=100 \mu s$ to model the hardware constraints, and the throttling parameter is set to $k=0.01$.
Unless otherwise specified, the simulations use the default parameters in~\cref{tab:simulation-parameters-table}.

\para{Methodology}
We evaluate \system{} against a standard DCQCN-enabled RoCEv2 baseline.
This mirrors the prevailing configuration in various large-scale production AI clusters~\cite{MegaScaleBytedance10000GPUsNSDI2024, AlibabaHPNSIGCOMM2024}. 
The metrics used for evaluations are Job Completion Time (JCT) and Collective Completion Time (CCT). 
All reported metrics (JCT and CCT) are averaged across at least 20 runs using random seeds.

\subsection{Effectiveness in mitigating step misalignment}
\label{subsubsec:reduce-misalignment}

\begin{figure}[t]
    \centering
    \begin{subfigure}{\linewidth}
        \centering
        \includegraphics[width=0.85\linewidth]{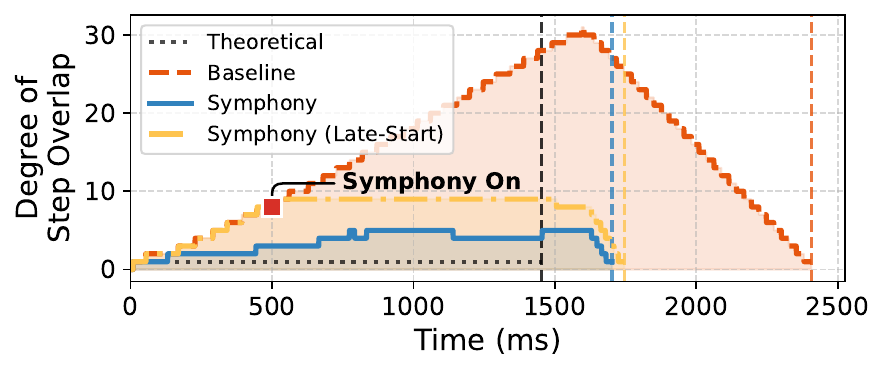}
        \vspace{-1em} 
        \caption{
        Degree of step overlap over time.
        }
        \label{fig:evaluation-reduce-misalignment-example}
    \end{subfigure}
    \hfill        
    \begin{subfigure}{\linewidth}
        \centering
        \includegraphics[width=0.75\linewidth]{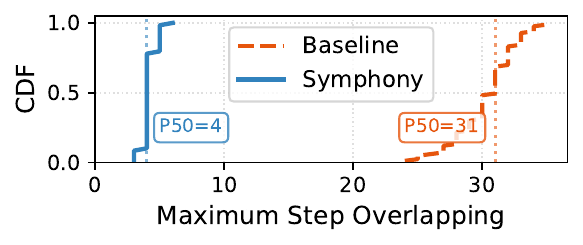}
        \vspace{-1em}
        \caption{CDF of the maximum step overlap.}
        \label{fig:evaluation-reduce-misalignment-cdf}
    \end{subfigure}
    \vspace{-1em} 
    \caption{Step misalignment mitigation effectiveness.}
    \label{fig:effectiveness-reduce-misalignment}
    \vspace{-1em} 
\end{figure}

We demonstrate that \system can effectively minimize step misalignment for ring-based collective operations.
We replicate~\cref{fig:motivation-example-experiments} and compare the performance of the Baseline against \system.
The results are depicted in~\cref{fig:effectiveness-reduce-misalignment}.

\para{Comparison with baseline}
For the Baseline case, \cref{fig:evaluation-reduce-misalignment-example} reproduces the ``snowball effect'' highlighted in \cref{subsec:ring-non-ideal}: under ECMP routing, minor timing skews compound into severe misalignment, with up to 30 overlapping steps and inflating CCT to 60\% more than the theoretical value.

As for \system{}, we observe that the number of overlapping steps is kept low, i.e., no more than 5 steps throughout the entire execution (see~\cref{fig:evaluation-reduce-misalignment-example}). 
Even with ECMP, this ``clamping'' effect directly limits the runaway misalignment and reduces CCT by about 30\% as compared to the Baseline.

We evaluate the scenario of \emph{\system (Late Start)}, where \system{} is only activated \ms{500} into the session.
Despite an initial accumulation of 10 overlapping steps, \system  prevents further divergence and keeps the number of overlapping steps under control (unlike Baseline), and effectively brings down the CCT.

\para{Robustness}
We further test the robustness of \system by observing its behavior over multiple runs. 
\Cref{fig:evaluation-reduce-misalignment-cdf} shows the CDF of maximum step overlap across 50 runs. 
While Baseline shows maximum step overlap between 24 and 35, \system{} consistently keeps the number of overlapping steps low, with maximum step overlap in the range between 3 and 6 over all runs.
This further substantiates that \system{} is indeed effective in mitigating step misalignments.

\subsection{{Collective communication performance}}
\label{subsec:collective-comm}

\begin{figure}
    \centering
    \includegraphics[width=0.75\linewidth]{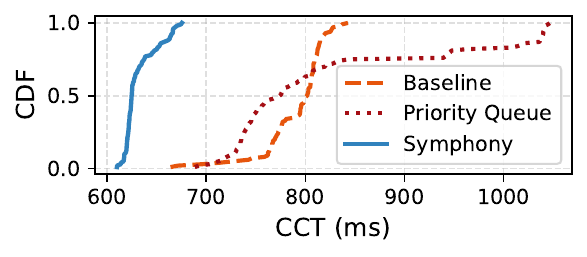}
    \vspace{-1em} 
    \caption{CDF of CCT for Ring AllReduce.}
    \label{fig:microbenchmark-cdf-different-solutions}
    \vspace{-1em} 
\end{figure}

We now evaluate the effectiveness of \system in reducing the CCT of collective operations. 
We analyze the distribution of CCT over 100 runs of a Ring AllReduce operation (following~\cref{tab:simulation-parameters-table}).
Additionally, we also compare \system{} against a \textit{priority queuing} (PQ) strategy (discussed in~\cref{subsec:baseline}) for lagging flows.
The results are shown in \cref{fig:microbenchmark-cdf-different-solutions}.

\para{Results}
We observe that \system{} consistently outperforms both baseline and PQ with substantially lower CCT, with $\approx$22\% and $\approx$19\% reduction at the median, respectively.

In the case of priority queuing, as it enforces strict priority for lagging flows, it inevitably causes starvation for non-prioritized flows, triggering aggressive DCQCN throttling, which explains the poor performance.
This explains why strict priority queuing is not suitable for addressing step misalignment, and hence highlights the importance of \system{}'s selective throttling approach to free up bandwidth for lagging flows to catch up and restore alignment.

\subsection{End-to-end AI training workloads}
\label{subsubsec:end-to-end}

\begin{table}
\centering
\caption{
    End-to-end data parallel test, gradient synchronization phase time comparison. 
    JCT is in ms.
}
\label{tab:end-to-end-data-parallel}
\resizebox{0.9\linewidth}{!}{%
\begin{tabular}
{p{0.21\linewidth}|p{0.17\linewidth}|p{0.17\linewidth}|l}
\hline
\textbf{Workload \newline \& Scale} & \textbf{Baseline \newline JCT} & \textbf{\system{} \newline JCT} & \textbf{Improvement} \\
\hline
\hline
VGG-128 & 2450.34 & 1220.22 & 50.2\% \\
VGG-512 & 2676.09 & 1220.30 & 54.4\% \\
ResNet-128 & 977.27 & 739.85 & 24.3\% \\
ResNet-512 & 1034.91 & 819.31 & 20.8\% \\
Transformer & 96389.86 & 96324.31 & 0.068\%\\
\hline
\end{tabular}
}
\vspace{-1em} 
\end{table}

\begin{figure}
    \centering
    \includegraphics[width=0.75\linewidth]{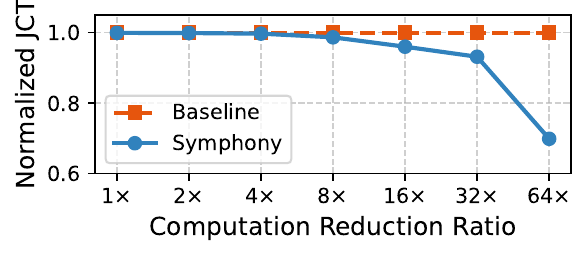}
    \vspace{-1em} 
    \caption{Normalized JCT for Transformer training under varying computation reduction ratios.}
    \label{fig:end-to-end-hybrid-speed-up}
    \vspace{-1em} 
\end{figure}

Next, we study how \system{} can improve AI model training workloads.
We evaluate three representative models: ResNet50~\cite{resnet} and VGG16~\cite{vgg} (Data Parallel, represents communication-bound workloads), and the original Transformer~\cite{attention-is-all-you-need} architecture (Hybrid Data \& Model parallelism, represents a compute-bound workload). 
These workloads are the common benchmarking suite for Astra-Sim~\cite{rashidi2020astrasim, won2023astrasim2}.
They cover a spectrum of communication patterns, from massive gradient synchronizations to small and frequent exchanges.

\para{Results}
From~\cref{tab:end-to-end-data-parallel}, we observe that \system achieves significant gains for communication-intensive workloads. 
For VGG16, the JCT is reduced by 50.2\% and 54.4\%, for job sizes of 128 and 512 nodes, respectively.
As for ResNet50, the reduction is 24.3\% and 20.8\%.
As expected, for the compute-bound Transformer baseline, there is a minimum difference since communication has a much smaller role.

To understand the impact of communication overhead, for the compute-bound Transformer baseline,~\cref{fig:end-to-end-hybrid-speed-up} shows how normalized JCT changes when the communication-to-computation ratio increases (to simulate the faster next-generation accelerators). 
The result shows that as normalized JCT decreases, reaching nearly 30\% when the relative computation times reduce by 64 times.

These results highlight the following.
\system{} is particularly effective for situations where collective communication constitutes a larger fraction of total job time, such as data-parallel workloads.
As \system{} prioritizes mitigating the impact of lagging flows, there are also more improvements in jobs where collective steps vary significantly in size or cost, such as VGG workloads with both small and large collectives.
Finally, when the ratio of communication over computation increases, \system{} becomes more effective.

\subsection{Multi-tenant environment workloads}
\label{subsubsec:multi-tenant}

\begin{figure}
    \centering
    \begin{subfigure}{\linewidth}
        \centering
        \includegraphics[width=\linewidth]{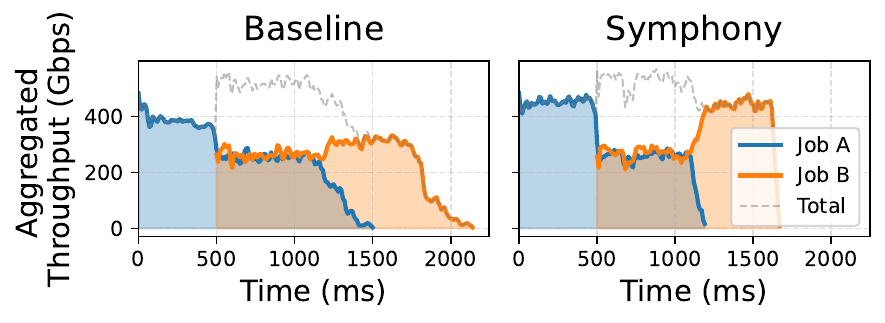}
        \vspace{-1em} 
        \caption{Throughput timeline of two co-located jobs.}
        \label{fig:multi-tenant-throughput-sharing}
    \end{subfigure}

    \begin{subfigure}{\linewidth}
        \centering
        \includegraphics[width=0.75\linewidth]{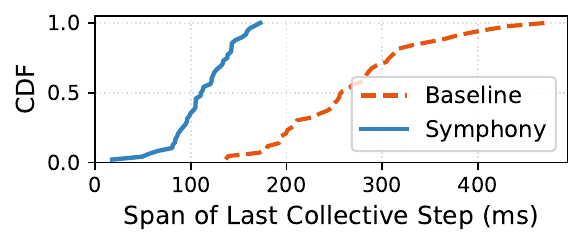}
        \vspace{-1em} 
        \caption{CDF of the time span of the final collective step.}
        \label{fig:multi-tenant-tail-skew}
    \end{subfigure}

    \begin{subfigure}{\linewidth}
         \centering
        \includegraphics[width=0.75\linewidth]{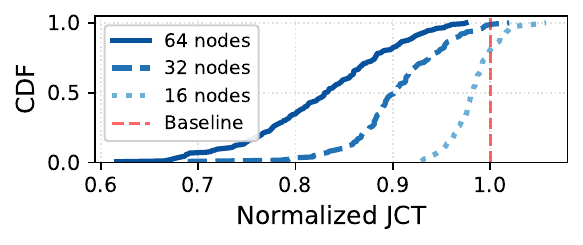}
        \vspace{-1em} 
        \caption{JCT improvement of \system across different job scales when multiple jobs are co-located with each other.}
        \label{fig:multi-tenant-jct-cdf}
    \end{subfigure}
    \vspace{-1em}
    \caption{
    Multi-tenant environment workloads 
    }
    \label{fig:multi-tenant-all}
    \vspace{-1em}
\end{figure}

Next, we evaluate Symphony's robustness under a multi-tenant environment.

\para{Two-job scenario}
We evaluate a co-location scenario where two identical jobs (job A followed by job B after \ms{500}) compete for the same bottleneck links on a 128-node cluster (configured as 32 parallel 1D rings, 16 rings each).

\cref{fig:multi-tenant-throughput-sharing} illustrates how throughput varies as the jobs overlap. 
Unlike the Baseline, where step misalignment degrades aggregate throughput, \system maintains a consistently higher throughput.
Note that \system eliminates the \linebreak ``heavy tail'' lagging flows seen in the Baseline, where Job A spends much more time in the ``completion phase'' when throughput starts to decrease as some flows have completed.  

In ~\cref{fig:multi-tenant-tail-skew}, we plot the CDF of the span of the final collective step, computed from the time difference between the completion times of the fastest and slowest flow over 50 runs. 
Clearly, \system consistently minimizes the span duration of the final collective step, minimizing the tail latencies of overlapping jobs.

\para{Multiple jobs with random arrival.}
We further simulated a dynamic environment on a 128-node cluster to evaluate robustness. 
To create dynamic resource contention, we generate a continuous stream of jobs (Ring AllReduce operations) with varying scales (16, 32, and 64 nodes), workload sizes (4, 16, and 32MB), and durations (controlled by varying the number of operation passes from 16 to 128).
The arrival times are randomized to simulate unpredictable job starts, maintaining a maximum concurrency of 3--4 jobs. 
The experiment is repeated for 50 runs with different random seeds. 

\cref{fig:multi-tenant-jct-cdf} shows the normalized JCT of \system's improvement over baseline.
We observe a clear trend where the benefits of \system amplify with job scale. 
For smaller 16-node jobs, the median JCT is reduced by 1.53\%, while the larger 64-node jobs achieve up to $\approx$17\% JCT reduction.
Note that for smaller jobs (16 nodes) where improvement is limited, \system does not degrade the performance as well. 
This validates \system as a safe ``performance insurance'': it delivers substantial gains for vulnerable large-scale jobs while retaining similar performance for lighter workloads.

\subsection{Impact of different network conditions and system parameters}
\label{subsubsec:robustness}

\begin{figure}[t]
    \centering
    \begin{subfigure}[t]{\linewidth}
        \centering
        \includegraphics[width=0.75\linewidth]{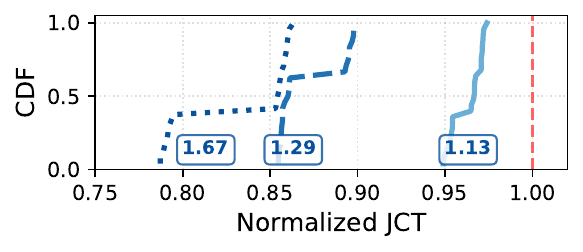}
        \caption{Impact of load imbalance ratios.} 
        \label{fig:load-imbalance-jct-cdf}
    \end{subfigure}
    \hfill
    \begin{subfigure}[t]{\linewidth}
        \centering
        \includegraphics[width=0.75\linewidth]
        {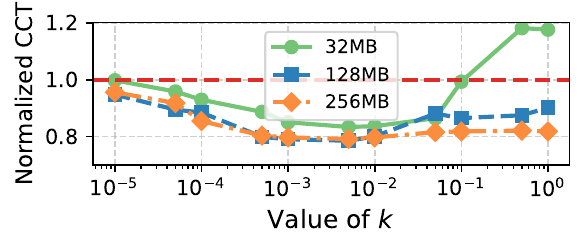}
        \caption{Impact of different $k$ values.}
        \label{fig:microbenchmark-k-value-study}
    \end{subfigure}
    \hfill
    \begin{subfigure}[t]{\linewidth}
        \centering
        \includegraphics[width=0.75\linewidth]
        {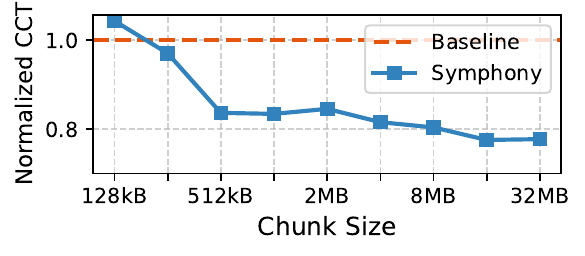}
        \caption{Impact of different chunk sizes.}
        \label{fig:microbenchmark_workload_size}
    \end{subfigure}
    \vspace{-1em}     
    \caption{
    Impact of different network conditions and system parameters.
    }
    \label{fig:sensitivity-analysis-all}
    \vspace{-1em} 
\end{figure}

Here, we study \system's behavior under varying network conditions, traffic patterns, and parameter configurations.
We employ the standard Multiple 1D Ring AllReduce to evaluate robustness against network imbalance under 128 nodes cluster.
For the impact of parameter $k$ and chunk size, we utilize a 2D Ring AllReduce pattern.

\para{Impact of load imbalance ratios}
Since load imbalance is a common driver of step misalignment (as discussed in~\cref{subsec:ring-non-ideal}), we evaluate \system's relation to load imbalance.
\cref{fig:load-imbalance-jct-cdf} plots the normalized JCT as we increase the network load imbalance ratio from 1.1x to 1.7x.
By including small load imbalance ratios 
(e.g., Meta reports an imbalance ratio of over 1.2 even in highly optimized clusters~\cite{MetaScaleSIGCOMM2024}),  we try to emulate state-of-the-art load balancing algorithms.
The results show a clear correlation between the amount of workload imbalance and \system's efficacy. 
As the imbalance increases, \system provides a larger (relative) improvement, and vice versa.
This shows that even with better load balancing algorithms, \system{} can consistently improve the CCT, and therefore highlights \system{}'s indispensable role even when good traffic optimization schemes are in place.

\para{Impact of different $k$ values}
Next, we evaluate how different values of $k$ (which controls the aggressiveness of \system{}'s throttling, see~\cref{subsec:straggler-mitigation}) impact performance\footnote{
We focus on the primary parameter $k$, as the architectural constants ($\tau, T_{win}$) are determined by the definition of misalignment (majority consensus) and switch processing latency.
}.
From the results shown in~\cref{fig:microbenchmark-k-value-study}, it indicates a broad ``sweet spot'' ($10^{-3}$ to $10^{-2}$) where performance remains consistently high. 
Performance degrades only at extreme values: $k \ge 0.1$ leads to over-reaction, whereas $k \le 10^{-4}$ yields insufficient feedback.
Crucially, this effective range spans an order of magnitude and, as observed in our experiments, remains consistent across varying flow sizes and traffic patterns.
This suggests that $k$ is insensitive to traffic characteristics and is instead determined by intrinsic network properties, such as RTT.
Therefore, in practice, operators can deploy \system with a single cluster-wide $k$ value, eliminating the need for fragile per-job parameter tuning.

\para{Impact of chunk size}
The chunk size impacts how much data is sent between nodes.
The longer the transmission duration, the more likely the flow is to be affected and thus making them more prone to step misalignments.
As shown in \cref{fig:microbenchmark_workload_size}, \system's gains are most pronounced with larger chunk sizes (i.e., $\ge$ \kbyte{512}), reducing the CCT relative to the baseline by up to $\approx$20\%.
Larger chunks create long-duration flows prone to compounding overlaps, which \system{} effectively mitigates.
Conversely, small chunks (e.g., \kbyte{128}) complete too quickly for significant misalignment to accumulate.

\subsection{Hardware prototype validation}
\label{subsec:testbed-eval}

\begin{figure}[t]
    \centering
    \includegraphics[width=\linewidth]{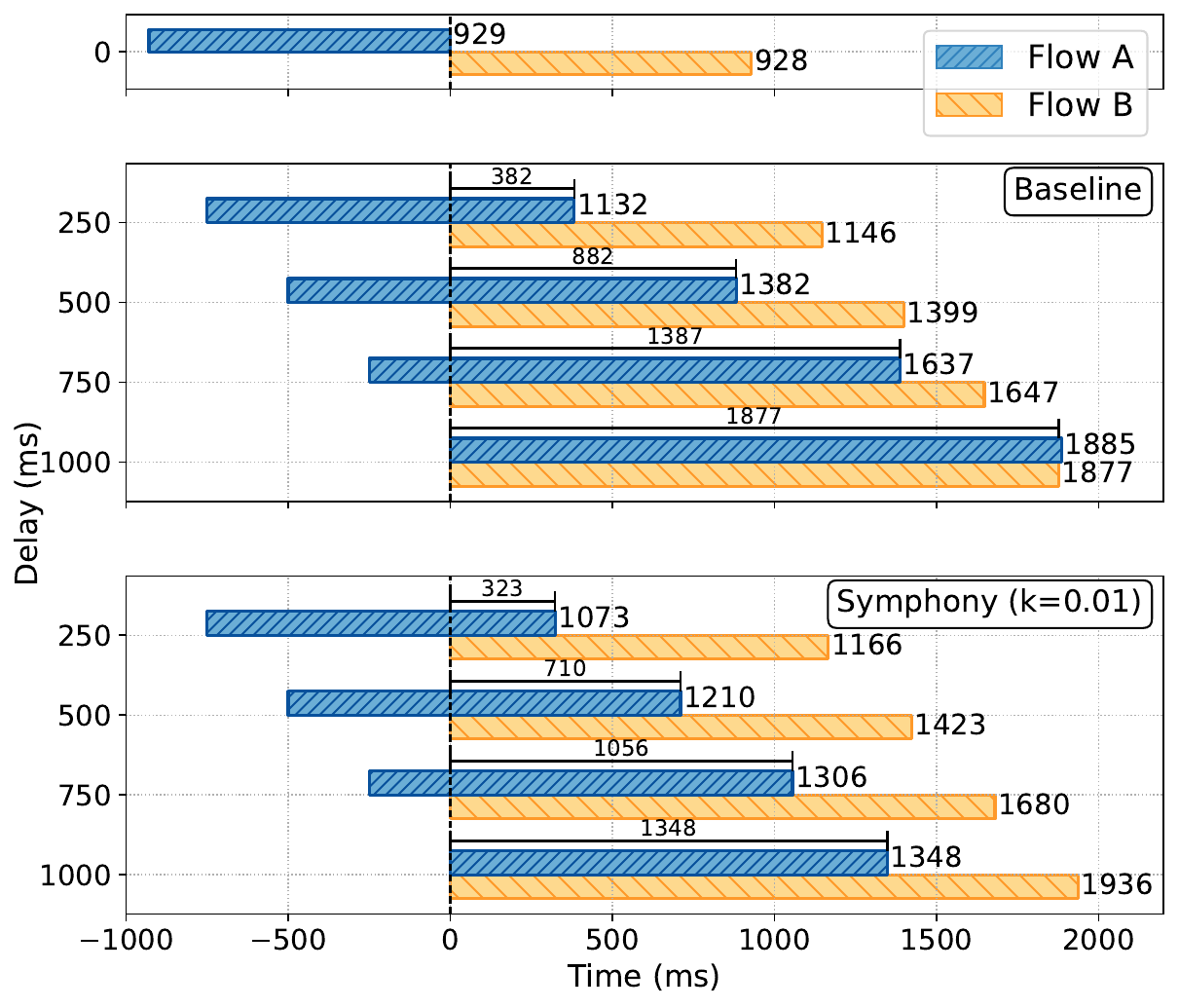}
    \caption{  
    HW prototype evaluation. 
    We show the transmission timelines of the data streams, and compare the ideal case (two flows run independently), the baseline (DCQCN-only), and \system{}.
    }
    \vspace{-1em}
    \label{fig:hardware-evaluation}
\end{figure}

To demonstrate the practicality of \system{}, we prototype \system{} on an Intel Tofino2~\cite{intel-tofino2} switch in about 1100 lines of P4~\cite{bosshart2014p4-long} code.

\para{Adapting to hardware constraints}
We utilize stateful ALUs to maintain the per-job states.
A key challenge in monitoring the outpacing traffic ratio (\cref{subsubsec:outpacing-traffic-ratio}) is that calculating $\rho(t)$ requires division, which is resource-intensive on switching ASICs. 
But as simplified by \cref{eqn:outpacing-check}, the multiplication-based inequality checking and additive operation make it workable on a switch. 
Similarly, calculating the precise ECN marking probability is non-trivial given that modern switching hardware only supports simple arithmetic operations~\cite{2021-NetSoft-dSketch}. 
We address this by approximating the marking probability using logarithms and hardware lookup tables.

\para{Experiment setup}
To evaluate our prototype, two flows (A and B) of the same size {(e.g., 1GB)} are transmitted through the switch and go through the same port. 
In our setup, DCQCN is enabled on our RDMA NICs (NVIDIA ConnectX-6) and the DCQCN parameters are set as per~\cite{2023-NSDI-AzureRDMA}.
Similar to the simulations, the NICs are set to 10G, and the application tags its packets with the step index in the UDP source port by using the \texttt{mlx5dv\_modify\_qp\_udp\_sport} API~\cite{rdma-core}.
We observe how long each flow takes to complete as we vary their starting times.
The results are shown in~\cref{fig:hardware-evaluation}.

\para{Results}
As a reference, we validate that if flow B starts immediately after flow A completes, without any overlap in transmission time, both flows complete in the same amount of time, with each flow fully utilizing the link when it is active.
As we delay the start time of flow A from \ms{250} (Flow A starts \ms{250} later than the baseline) to \ms{1000} (two flows start at the same time), we observe that the differences in CCT between baseline and \system increase as the amount of time the two flows transmit concurrently increases.

Compared to the baseline, for the same amount of time that flow A is being delayed, \system reduces the completion time of flow A significantly. 
At the same time, note that \system is also reducing the duration that both flows are transmitting concurrently, mitigating the misalignment in this step.
For example, when flow A starts \ms{500} later, completion time of flow A reduces from \ms{1382} to \ms{1210} (\percent{12} reduction), with a relatively small increase in completion time of flow B, from \ms{1399} to \ms{1433} (\percent{2} increase). 
The concurrent transmission duration between flow A and B also reduces from \ms{882} to \ms{710}. 

%% file: sections/discussion.tex
\section{Discussion and Future Work}
\label{sec:discussion}

\para{Practical deployment}
As \system{} relies on existing hardware capabilities (i.e., simple state tracking and ECN marking), it can be readily implemented on today's network hardware through a software/firmware update.
Additionally, as the selective throttling mechanism depends on identifying progress at the ingress, \system{} only needs to be deployed on ToR switches. 
Combined, they eliminate the need for extensive changes as well as hardware upgrades to support \system{}.
At the same time, \system{} can be deployed incrementally across the cluster, given how each switches perform flow tracking and selective throttling independently without coordination.
This allows operators to gradually introduce \system{} in their network in a controlled manner.

\para{Large-scale testbed evaluations}
Given the lack of access to large-scale testbed infrastructure, our current evaluation focuses on using packet-level simulation and hardware prototype to validate the mechanism and feasibility of \system{}. 
To better quantify the gains of \system{} under more realistic network conditions, we plan to explore the deployment of \system{} in large-scale testbeds as part of future work.
Considering the massive scale of LLM training, where even a \percent{0.01} efficiency gain for a model like Llama3-405B saves over 10K GPU hours~\cite{dubey2024llama3herdmodels}, quantifying these benefits in a real-world cluster is our priority for future work.

\para{General applicability of \system{}}
By design, \system{} caters only to workloads dependent on ring-based collective operations.
While the evaluation of \system{} on emerging Mixture‑of‑Experts (MoE) workloads was absent in our simulations, given framework limitations, it is important to note that ring‑based collectives remain the dominant communication primitive in dense LLM training and ZeRO‑style sharding~\cite{ZeRO-2020-arxiv, dubey2024llama3herdmodels}.
As such, this highlights the importance of \system{}.
Beyond ring-based collectives, we plan to explore how \system{} can be adapted to different communication patterns (e.g., tree) in future work.

\para{Practical concerns}
Unlike strict priority scheduling or admission control schemes that risk starvation, \system{} imposes a \textit{soft limit} via probabilistic ECN marking.
Crucially, this pacing mechanism is inherently bounded by the underlying transport stack (e.g., DCQCN~\cite{2015-SIGCOMM-DCQCN}), which enforces minimum sending rates even under aggressive marking.
In scenarios of internal failure, such as control plane timeouts or state loss, the selective throttling of \system{} simply ceases to mark any flows. 
Consequently, the network falls back to the baseline behavior, without compromising network connectivity.
Additionally, \system{} operates transparently to non-ring-based operations, as it only applies tracking and throttling on ring-based collective operations. 

%% file: sections/related_work.tex
\section{Related Work}
\label{sec:related_work}

\para{Collective communication optimization}
There have been various attempts to enhance the efficiency and scalability of collective communications.
Examples include routing and traffic engineering by leveraging datacenter topology~\cite{ncclrailoptimized, wang2024railonlylowcosthighperformancenetwork, AlibabaHPNSIGCOMM2024}, traffic characteristics~\cite{2024-SIGCOMM-TECCL, 2024-SIGCOMM-MCCS, 2025-SIGCOMM-ResCCL}, traffic aggregation~\cite{2021-NSDI-SwitchML, 2021-NSDI-ATP, 2023-ASPLOS-INA}, and even specialized hardware like optical switches~\cite{2025-SIGCOMM-MixNet}.
Others focus on synthesizing efficient collective algorithms~\cite{2025-TACO-IBing} or automating parallelism strategies~\cite{alpaOSDI2022, 2023-NSDI-TACCL} to maximize bandwidth utilization.
Recent works also propose advanced scheduling mechanisms to prioritize critical flows or mitigate pipeline bubbles~\cite{2024-SIGCOMM-CRUX, 2026-NSDI-straggler}.
In parallel, payload-centric methods exploit ML traffic characteristics to reduce communication overhead through compression and fusion~\cite{2024-SC-gZCCL, 2025-HPDC-SAFusion, MLSYS-2023-cupcake-compression}, or employ data redundancy to handle heterogeneity and stragglers~\cite{2019-NIPS-redundancy, 2025-NSDI-optireduce, 2024-SIGCOMM-ccl-consumer-gradegpu}.
However, these works largely treat collective operations as a black box or focus solely on data volume reduction.
They do not examine the fine-grained temporal synchronization of the collective operation itself.
While diagnosis tools like Mycroft~\cite{2025-SOSP-Mycroft} improve observability by tracing dependencies to debug reliability issues, they are not designed to mitigate the misalignment problem actively.
\system{} examines the black box to target fine-grained progression within collective operations, identifies \textit{step misalignment} as a critical inefficiency, and introduces a mechanism to mitigate its impact.

\para{Congestion signaling and marking}
To alleviate congestion in datacenter networks, ECN has been widely deployed as a key mechanism for achieving high throughput and low latency.
Existing ECN marking strategies can be broadly categorized into three paradigms.
First, \textit{per-port marking} applies a uniform threshold to all packets traversing a physical port, deriving from the classic RED algorithm~\cite{RandomEarlyDetectionRED, ECN-rfc3168, yan2021acc, bai2016enabling, zhang2019enabling, 2015-SIGCOMM-DCQCN}.
Second, \textit{per-queue marking} sets distinct thresholds for different output queues or service classes to provide isolation~\cite{nichols2012controlling, bai2016nsdi, 2015-NSDI-Hull}.
Third, \textit{per-flow marking} assigns differentiated thresholds or drop probabilities at the granularity of individual flows to ensure fairness or minimize flow completion times~\cite{hoeiland2018flow, DiffECN, sharma2018approximating}.
\system{} falls under a flow-level differentiated strategy but introduces a critical distinction.
Unlike prior approaches that differentiate flows based on static attributes (e.g., flow size) or fairness metrics, \system{} adopts a \textit{progress-aware} selective scheme. 
By marking outpacing flows while suppressing signals for lagging ones, it actively enforces temporal synchronization for collective operations.

\para{Scheduling-based queue management} 
Scheduling algorithms differentiate packet forwarding to satisfy diverse QoS objectives.
Classic mechanisms enforce bandwidth fairness~\cite{demers1990analysis, shreedhar1996efficient, goyal1996start}, 
minimizing FCT~\cite{schrage1966queue, alizadeh2013pfabric}, 
ensuring deadline guarantees~\cite{EDF}, or focus on minimizing CCT~\cite{varys, Sincronia, D-CLAS, Philae}. 
While the latter is particularly important for collective communication, these approaches primarily manage \textit{inter-job} contention and fall short in addressing the \textit{intra-job} step misalignment identified in this paper.
Moreover, step misalignment can unpredictably prolong job processing time, distorting the completion time estimates these algorithms rely on, leading to incorrect scheduling decisions.

%% file: sections/conclusion.tex
\section{Conclusion}
In this paper, we identify step misalignment as a fundamental yet underexplored bottleneck in ring-based collective communication, arising from subtle progress divergence across flows under realistic datacenter conditions.
We reveal how steps overlap and compound over time, significantly inflating collective completion time.
Motivated by this insight, we explore an in-network approach \system{} that can efficiently detect and mitigate step misalignments entirely in the data plane.
Our findings highlight the importance of aligning collective operations at the step granularity and point to new directions in co-designing network mechanisms with collective communications.

\textbf{This work does not raise any ethical concerns.}